\documentclass[sigconf]{acmart}

\usepackage{array}
\usepackage{graphicx}
\usepackage{clrscode}
\usepackage{balance}
\usepackage{subfigure}
\usepackage{multirow}
\usepackage{booktabs}
\usepackage{adjustbox}
\usepackage[justification=justified,skip=5pt]{caption}
\usepackage{float}
\usepackage{color}
\usepackage{soul}
\usepackage{mdframed}
\usepackage{amsopn}
\usepackage{mathrsfs}
\usepackage{mathtools}
\usepackage{amsmath}
\usepackage{arydshln}
\usepackage{hyperref}
\usepackage{multicol}
\usepackage{blkarray}
\usepackage{enumerate}
\usepackage{courier}
\usepackage{rotating}
\usepackage{booktabs}
\usepackage{diagbox}
\usepackage{fancybox}
\usepackage{minibox}
\usepackage{cases}
\usepackage{bm}
\usepackage{enumitem}
\usepackage[lined, ruled, commentsnumbered, linesnumbered]{algorithm2e} %
\usepackage{enumitem}
\setlist[itemize]{leftmargin=*}

\AtBeginDocument{%
  \providecommand\BibTeX{{%
    \normalfont B\kern-0.5em{\scshape i\kern-0.25em b}\kern-0.8em\TeX}}}

\copyrightyear{2022}
\acmYear{2022}
\setcopyright{acmcopyright}
\acmConference[SIGIR '22] {Proceedings of the 45th International ACM SIGIR Conference on Research and Development in Information Retrieval}{July 11--15, 2022}{Madrid, Spain.}
\acmBooktitle{Proceedings of the 45th International ACM SIGIR Conference on Research and Development in Information Retrieval (SIGIR '22), July 11--15, 2022, Madrid, Spain}
\acmPrice{15.00}
\acmISBN{978-1-4503-8732-3/22/07}
\acmDOI{10.1145/3477495.3531973}

\begin{document}
\fancyhead{}

\title[Explainable Fairness in Recommendation]{Explainable Fairness in Recommendation}



\settopmatter{authorsperrow=5}

\author{Yingqiang Ge}
\affiliation{\normalsize
  \institution{Rutgers University}
}
\email{yingqiang.ge@rutgers.edu}

\author{Juntao Tan}
\affiliation{\normalsize
  \institution{Rutgers University}
}
\email{juntao.tan@rutgers.edu}

\author{Yan Zhu}
\affiliation{\normalsize
  \institution{Meta Platforms, Inc.}
}
\email{yzhu@fb.com}

\author{Yinglong Xia}
\affiliation{\normalsize
  \institution{Meta Platforms, Inc.}
}
\email{yxia@fb.com}

\author{Jiebo Luo}
\affiliation{\normalsize
  \institution{University of Rochester}
}
\email{jluo@cs.rochester.edu}

\author{Shuchang Liu}
\affiliation{\normalsize
  \institution{Rutgers University}
}
\email{shuchang.liu@rutgers.edu}

\author{Zuohui Fu}
\affiliation{\normalsize
  \institution{Rutgers University}
}
\email{zuohui.fu@rutgers.edu}

\author{Shijie Geng}
\affiliation{\normalsize
  \institution{Rutgers University}
}
\email{shijie.geng@rutgers.edu}

\author{Zelong Li}
\affiliation{\normalsize
  \institution{Rutgers University}
}
\email{zelong.li@rutgers.edu}

\author{Yongfeng Zhang}
\affiliation{\normalsize
  \institution{Rutgers University}
}
\email{yongfeng.zhang@rutgers.edu}


\renewcommand{\authors}{Yingqiang Ge, Juntao Tan, Yan Zhu, Yinglong Xia, Jiebo Luo, Shuchang Liu, Zuohui Fu, Shijie Geng, Zelong Li, Yongfeng Zhang}
\begin{abstract}
Existing research on fairness-aware recommendation has mainly focused on the quantification of fairness and the development of fair recommendation models, neither of which studies a more substantial problem--identifying the underlying reason of model disparity in recommendation. This information is critical for recommender system designers to understand the intrinsic recommendation mechanism and provides insights on how to improve model fairness to decision makers. Fortunately, with the rapid development of Explainable AI, we can use model explainability to gain insights into model (un)fairness. In this paper, we study the problem of \textit{explainable fairness}, which helps to gain insights about why a system is fair or unfair, and guides the design of fair recommender systems with a more informed and unified methodology. 
Particularly, we focus on a common setting with feature-aware recommendation and exposure unfairness, but the proposed explainable fairness framework is general and can be applied to other recommendation settings and fairness definitions. We propose a Counterfactual Explainable Fairness framework, called CEF, which generates explanations about model fairness that can improve the fairness without significantly hurting the performance. The CEF framework formulates an optimization problem to learn the ``minimal'' change of the input features that changes the recommendation results to a certain level of fairness.
Based on the counterfactual recommendation result of each feature, we calculate an explainability score in terms of the fairness-utility trade-off to rank all the feature-based explanations, and select the top ones as fairness explanations. Experimental results on several real-world datasets validate that our method is able to effectively provide explanations to the model disparities and these explanations can achieve better fairness-utility trade-off when using them for recommendation than all the baselines.
\end{abstract}





\maketitle

\section{Introduction}
Nowadays, with the extensive deployment in various e-commerce platforms, recommender systems (RS) have been widely acknowledged for their strong capabilities of delivering high-quality services to users~\cite{ ge2020learning,xu2021causal,liu2021variation,fu2021hoops,ge2020understanding}.
Despite these huge benefits, the issue of fairness in recommendation has also attracted considerable interests from both academia and industry~\cite{Geyik2019,singh2018fairness,li2021cikm,lin2017fairness}.
Fortunately, these concerns about algorithmic fairness have resulted in a resurgence of interest to develop fairness-aware recommendation models to ensure that such models would not become a source of unfair discrimination in recommendation \cite{Mehrotra2018,ekstrand2019fairness,burke18a,zhu2018fairness}.
In the area of fairness-aware recommendation, existing research mainly focus on the quantification of fairness and the development of fair recommendation models.
Fairness quantification aims to develop and investigate quantitative metrics that measure algorithmic disparities in ranking or recommendation \cite{fu2020fairness,gao2021fair,li2021user}.
For example, \cite{fu2020fairness,li2021user} proposed and studied the recommendation quality unfairness between active users and inactive users.
Meanwhile, fair recommendation aims to find feasible algorithmic approaches that can adjust the recommendation results to reduce recommendation disparities.
For example, \cite{ge2021towards, abdollahpouri2019unfairness} proposed approaches to mitigating the popularity bias between different item groups.

Despite the great efforts on fairness-aware recommendation and possibly countless future emergence of discoveries, one fundamental question that has not been studied extensively yet is fairness diagnostics, i.e., 
\begin{itemize}
    \item \textbf{RQ} \textit{What are the sources that result in model disparities in recommendation?}
\end{itemize}
Considering the huge commercial and social values that recommender systems bring to various web platforms and the society, we believe that the answer to this \textbf{RQ} is critical for recommendation system designers to understand the intrinsic recommendation mechanism and to provide insights for decision makers on how to improve model fairness.
Yet, the answer to this question turns out to be unsurprisingly challenging especially when the predictive model is a large-scale deep black-box model with large numbers of input features.
For example, it is hard to tell how input features (such as screen size, battery, camera) would influence the exposure unfairness.
Note that some pioneer works in other areas have leveraged Explainable AI to seek for feature-based explanations for certain fairness outcome.
For instance, \citeauthor{begley2020explainability} used Shapley value to attribute the model disparity in classification \cite{begley2020explainability,pan2021explaining}.
Though their methods successfully provide explanations to the model disparities in simple tasks, they are not suitable for recommender systems where the model inputs could be extremely large and sparse, which may bring huge computational cost when calculating Shapley value for each feature.
Furthermore, existing methods only partially answers the above question, since they only explain either utility or fairness alone, ignoring the fact that there is an inherent trade-off between fairness and utility, which has been demonstrated by several recent work both empirically and theoretically~\cite{ge2022toward,lin2017fairness,kamani2021pareto,kearns2019ethical,NEURIPS2018_8e038477,10.5555/3294771.3294793}.
And this incomplete view may potentially downgrade the stringency of the method because explanations that have the same effect on model fairness may not have the same effect on model utility.

In this paper, we propose a novel framework to explain the recommendation (un)fairness based on a counterfactual reasoning paradigm.
Particularly, we focus on a common setting with feature-aware recommendation and item exposure unfairness (popularity bias) \cite{ge2021towards,abdollahpouri2019unfairness,abdollahpouri2017controlling} since feature-base explanations are more straightforward and easy to understand, which would be a great demonstration of the effectiveness of our method. 
However, the proposed approach is very general and can be applied to other recommendation settings with various fairness definitions.
Specifically, we propose a Counterfactual Explainable Fairness (CEF) framework to generate feature-based explanations in terms of item exposure disparity for various black-box feature-aware recommendation models.
We first follow prior works \cite{zhangsigir14,chen2020try,Wang2018} to build a user-feature attention matrix as well as an item-feature quality matrix, and use both matrices to train a feature-aware recommendation model.
Then, we aim to find the ``minimal'' changes to a given feature in the feature space that switch the recommendation results to a certain level of fairness.
To avoid overwhelmed sacrificing of the recommendation quality, we also constrain the feature perturbation within a certain degree in the objective function.
With the counterfactual learning objective and the perturbation constraint, our proposed framework is able to generate feature-level explanations that consider the fairness-utility trade-off.
Finally, we calculate an explainability score in term of the fairness-utility trade-off based on the counterfactual recommendation result of each feature.
These scores help rank the feature-based explanations and we select the top ones as fairness explanations for the pre-trained recommendation model. 

In general, the contributions of this work can be summarized as follows:
\begin{itemize}
    \item We study the problem of explainable fairness in recommendation and propose a framework based on counterfactual reasoning. 
    To the best of our knowledge, this is the first work that introduces explainable fairness in recommender systems.
    \item We design a learning-based intervention method to discover critical features that will significantly influence the fairness-utility trade-off and use them as fairness explanations for black-box recommendation systems;
    \item We conduct extensive experiments to evaluate our framework's effectiveness and validate that explanations generated by CEF can achieve better fairness-utility trade-off when using them for recommendation than all the baselines.
\end{itemize}

\section{Related Work}
There are several main research lines related to our work: explainable recommendation, fairness in recommendation and fairness explanation. We will briefly introduce each of them in this section.

\subsection{Explainable Recommendation}
Explainable recommendation has been an important topic in both academia and industry,
which helps to improve the transparency, user satisfaction and trust over the recommender systems \cite{Zhang2020,zhangsigir14}.
Early approaches mainly attempt to make latent factor models explainable by aligning each latent fator with an explicit meaning such as item features \cite{zhangsigir14,zhangsentiment14,chen2016learning}.
Recently, with the ever prospering of deep learning technology, many neural algorithms are developed to explain recommendations based on neural models.
For example, \cite{seo2017interpretable} proposed to attentively highlight particular words in user reviews as explanations, \cite{li2021extra,chen2019dynamic} proposed to rank user review sentences as explanations,
\cite{chen2019personalized,geng2022improving} proposed visually explainable recommendation to highlight image regions or directly generate image as explanations,  \cite{li2021personalized,li2020generate,chen2019generate,geng2022recommendation,geng2022improving,li2022personalized} proposed to generate natural language explanations, \cite{balog2019transparent} proposed set-based explanation for scrutability, \cite{xian2021ex3} proposed contrastive explanations for comparison shopping, and \cite{zhu2021faithfully,chen2021neural,chen2022graph,xian2020cafe,shi2020neural} proposed neural-symbolic methods to improve both explainability and accuracy.
In addition to text-based or image-based explainable recommendation, knowledge-aware explainable recommendation has also attracted research attention recently, such as \cite{wang2018ripplenet,ai2018learning,xian2019reinforcement,xian2020cafe,fu2020fairness}.



Works using counterfactual reasoning to improve recommendation explainability \cite{Ghazimatinwsdm2020,xu2021learning,tran2021counterfactual,tan2021counterfactual,tan2022learning} have been proposed very recently.
\citeauthor{Ghazimatinwsdm2020} \cite{Ghazimatinwsdm2020} tried to generate provider-side counterfactual explanations by looking for a minimal set of user's historical actions (e.g. reviewing, purchasing, rating) such that the recommendation can be changed by removing the selected actions.
\citeauthor{xu2021learning} \cite{xu2021learning} proposed to improve this by using perturbation model to obtain counterfactuals.
\citeauthor{tran2021counterfactual} \cite{tran2021counterfactual} adopted influence functions for identifying training points most relevant to a recommendation while deducing a counterfactual set for explanations.
\citeauthor{tan2021counterfactual} \cite{tan2021counterfactual} proposed to generate and evaluate explanations that considers the causal relations to the outcome.

Yet, our work is different from prior works on two key points:
1) In terms of problem definition, prior works generate counterfactual explanations to explain user behaviors or recommendation results, while our method generates such explanations to explain the fairness-utility trade-off in recommendation. 
2) In terms of technique, our method adopts a counterfactual reasoning framework from a global perspective, which explains the entire model behavior, while prior works focus on generating individual explanations for an individual recommendation result.

\subsection{Fairness in Recommendation}
The issue of fairness in recommendation has received growing concerns as recommender systems touch and influence people's daily lives more deeply and profoundly \cite{li2021towards,ge2022toward,wu2022joint}.
Several recent works focusing on fairness quantification have found various types of bias and unfairness in recommendations, such as gender and race~\cite{chen2018investigating, li2021towards, Yao2017}, item popularity~\cite{ge2021towards,abdollahpouri2019unfairness, abdollahpouri2017controlling,ge2022toward}, and user activeness~\cite{fu2020fairness,li2021user}. 
Meanwhile, the relevant methods for fair recommendation focusing on providing fair recommendation results based on pre-defined fairness, can be roughly divided into three categories: pre-processing, in-processing and post-processing algorithms \cite{li2021cikm}.
First of all, pre-processing methods usually aim to minimize the bias in the data sources.
It includes fairness-aware sampling methodologies in the data collection process to cover items of all groups, balancing methodologies to increase coverage of minority groups, and repairing methodologies to ensure label correctness \cite{10.1145/3404835.3462807}.
Secondly, in-processing methods aim at encoding fairness as part of the objective function, typically as a regularizer \cite{abdollahpouri2017controlling,beutel2019fairness,ge2021towards,li2021towards}.
Finally, post-processing methods modify the presentation of the results, e.g., by re-ranking through linear programming \cite{li2021user,singh2018fairness,yang2021maximizing} or multi-armed bandit \cite{celis2019controlling}.
Based on the characteristics of the recommender system itself, there also have been a few works related to multi-sided fairness in multi-stakeholder systems ~\cite{burke18a,Gao2019how}. 

Moreover, there are two primary paradigms adopted in recent studies on algorithmic discrimination: individual fairness and group fairness \cite{li2021cikm}: individual fairness requires that each similar individual should be treated similarly; and group fairness requires that the protected groups should be treated similarly to the advantaged group or the populations as a whole.
In this paper, we mainly focus on the item popularity fairness, which is a kind of group fairness and aims to achieve fair chances of exposure for different item groups \cite{ge2021towards,abdollahpouri2019unfairness,abdollahpouri2017controlling}.

\subsection{Fairness Explanation}
Explainability and fairness are two important perspectives for responsible recommender systems, however, the relationship between the two is still less explored.
There have been several pioneering studies trying to derive explanations for model fairness \cite{begley2020explainability,pan2021explaining} in other tasks.
For example, \citeauthor{begley2020explainability} \cite{begley2020explainability} leveraged Shapley value paradigm \cite{shapley201617} to attribute the feature contributions to model disparity to generate explanations.
It estimates the sum of individual contributions from input features, so as to understand which feature contributes more to the model disparity \cite{begley2020explainability}.
Though this type of methods successfully provide explanations to the model disparities, they are not suitable for recommender systems.
First of all, the definition of Shapley value is the average marginal contribution of a feature value across all possible coalitions, meaning that the computation time increases super-exponentially with the number of features.
In recommendation systems, this becomes impractical since it is very common to have a large number of user/item features in the feature space.
Secondly, the Shapley value can only explain either utility or fairness alone \cite{begley2020explainability,pan2021explaining}, but not the fairness-utility trade-off. 
However, our proposed Counterfactual Explainable Fairness (CEF) framework is able to mitigate the above problems.


\section{Explainable Fairness}
In this section, we first introduce how to use review information to generate user-feature matrix and item-feature matrix, then introduce the details of feature-aware recommendation systems.
We introduce how to generate counterfactual explanations for fairness in section \ref{sec:cef} and \ref{sec:es}.

\subsection{Feature Generation}
Suppose we have a user set with $m$ users denoted as $\mathcal{U}$, an item set $\mathcal{V}$ with $n$ items and their interaction set $\mathcal{T}=\{(u,v)|u \in \mathcal{U}, v \in \mathcal{V},  u \text{ has interacted with } v\}$.
Based on an open source toolkit for phrase-level sentiment analysis, called ``Sentires''\footnote{https://github.com/evison/Sentires}, we can easily convert the raw review information into a set of quadruples $\mathcal{W}=\{(u_l,v_l,f_l,s_l)\}^N_{l=1}$.
Specially, each element $(u_l,v_l,f_l,s_l) \in \mathcal{W}$ means user $u_l \in \mathcal{U}$ mentioned feature $f_l \in \mathcal{F}$ of item $v_l \in \mathcal{V}$ with sentiment $s_l \in \mathcal{S}$, where $\mathcal{F}$ denotes the set of all features with size $r$ and the sentiment set $\mathcal{S} = \{positive(+1),negative(-1)\}$. 
For example, in the review of “I like the color of this sweater, but the sleeve is not satisfied, since it is too tight for me.”, the features are “collar” and “sleeve”, and the user expresses positive and negative sentiments on them. The final extracted tuples are “(user, item, color,
positive)” and “(user, item, sleeve, negative)”, respectively.
Following the same method described in \cite{zhangsigir14,chen2020try,tan2021counterfactual}, we construct a user-feature attention matrix $\boldsymbol{A}\in\mathbb{R}^{m\times r}$ and an item-feature quality matrix $\boldsymbol{B} \in \mathbb{R}^{n\times r}$ using all the quadruples in $\mathcal{W}$, where $\boldsymbol{A}_{u, f}$ indicates to what extent the user $u$ cares about the feature $f$, and $\boldsymbol{B}_{v, f}$ indicates how well the item $v$ performs on the feature $f$. 
Specifically, $\boldsymbol{A}$ and $\boldsymbol{B}$ are calculated as:
\begin{equation}
\begin{aligned}
\label{eq:XY}
    \boldsymbol{A}_{u,f} &=
    \begin{cases}
    0,~\text{if user}~u~\text{did not mention feature}~f\\
    1+(M-1)\Big(\frac{2}{1+\exp(-t_{u,f})}-1\Big),~\text{else}
    \end{cases}\\
    \boldsymbol{B}_{v,f} &=
    \begin{cases}
    0,~\text{if item}~v~\text{has no review on feature}~f\\
    1+\frac{M-1}{1+\exp(-t_{v,f}\cdot \Bar{t}_{v,f})},~\text{else}
    \end{cases}
\end{aligned}
\end{equation}
where $M$ is the rating scale in the system, which equals to 5 (stars) in most cases, $t_{u,f}$ is the frequency that user $u$ mentioned aspect $f$, $t_{v,f}$ is the frequency that item $v$ is mentioned on feature $f$, and $\Bar{t}_{v,f}$ is the average sentiment of these mentions. 
For both $\boldsymbol{A}$ and $\boldsymbol{B}$ matrices, their elements are re-scaled into the range of $(1,M)$ using the sigmoid function (see Eq.\eqref{eq:XY}) to match with the original system's rating scale.
Readers may refer to \cite{zhangsentiment14, zhangsigir14} for more details and the same user-feature and item-feature matrix construction technique can also be found in \cite{Wang2018,gao2019explainable,le2021explainable,tan2021counterfactual}.

\subsection{\textbf{Feature-aware Recommender Systems}}
Once given the user-feature attention matrix $\boldsymbol{A}$ and item-feature quality matrix $\boldsymbol{B}$, we define a ranking model $g$ that predicts the user-item ranking score $\hat{y}_{i,j}$ for user $u_i$ and \mbox{item $v_j$ by:}
\begin{equation}
    \hat{y}_{u,v} = g(\boldsymbol{A}_u, \boldsymbol{B}_v\mid Z, \Theta)
\end{equation}
where $\boldsymbol{A}_u$ and $\boldsymbol{B}_v$ are the vector of user $u$ and the vector of item $v$, $\Theta$ is the model parameter, and $Z$ represents all other auxiliary information. 
Depending on the application, $Z$ could be rating scores, clicks, text, images, etc., and is optional in the recommendation model $g$.

In this work, we explore different implementations of $g$ to demonstrate the effectiveness of our proposed framework. 
The general architecture of $g$ is a multi-layer neural network, that is:
\begin{equation}
\small
    \hat{y}_{u v}=\boldsymbol{W}_{T} \sigma_{T}\left(\ldots\left(\boldsymbol{W}_{1} \sigma_{1}\left(merge\left(\boldsymbol{A}_{u}, \boldsymbol{B}_{v}\right)\right)+\boldsymbol{b}_{1}\right)+\ldots\right)+\boldsymbol{b}_{T}
\end{equation}
where, for the $t$-th layer $(t = 1, 2, \cdots, T), \sigma_{t}$ is a non-linear activation function, $\boldsymbol{W}_{t}$ and $\boldsymbol{b}_{t}$ are weights and bias terms, respectively. 
$merge(\cdot)$ is an operator merging the user-feature and item-feature vectors, and we explore it within the following functions:

\begin{itemize}
    \item \textbf{Element-wise Product Merge}:
\begin{equation}\label{eq:product}
merge\left(\boldsymbol{A}_{u}, \boldsymbol{B}_{v}\right)=\boldsymbol{W}_{U} \boldsymbol{A}_{u}^{T} \odot \boldsymbol{W}_{V} \boldsymbol{B}_{v}^{T}.
\end{equation}
where $\boldsymbol{W}_{U}$ and $\boldsymbol{W}_{V}$ are trainable parameters, and $\odot$ represents the element-wise product (a.k.a. Hadamard product).

    \item \textbf{Concatenation Merge}:
\begin{equation}\label{eq:concatenation}
    merge\left(\boldsymbol{A}_{u}, \boldsymbol{B}_{v}\right)=[\boldsymbol{W}_{U} \boldsymbol{A}_{u}^{T},\boldsymbol{W}_{V} \boldsymbol{B}_{v}^{T}].
\end{equation}
\end{itemize}


where $\boldsymbol{W}_U$ and $\boldsymbol{W}_{V}$ are trainable parameters.

Then, we train the model with a cross-entropy loss:
\begin{align}\label{eqn: train base}
\begin{aligned}
    Loss &= -\sum\limits_{u,v, y_{u,v}=1}\log \hat{y}_{u,v}-\sum\limits_{u,v, y_{u,v}=0}\log (1-\hat{y}_{u,v})\\
    &= -\sum\limits_{u,v} y_{u,v}\log \hat{y}_{u,v} + (1-y_{u,v})\log(1-\hat{y}_{u,v})
\end{aligned}
\end{align}
where $y_{u,v}=1$ if user $u$ previously interacted with item $v$, otherwise $y_{u,v}=0$. 

Generally, the recommendation model $g$ can be any ranking model as long as it takes the user-feature and the item-feature vectors as the input.
The implementation and training of $g$ will be detailed in the experiment section.


Finally, given $\{\mathcal{U},\mathcal{V},\mathcal{T},\boldsymbol{A},\boldsymbol{B},g\}$, our task is to generate feature-based explanations in terms of recommendation disparity for the black-box recommendation model $g$.
Besides, most of the important symbols used in the paper can be referred in Tab. \ref{Table:notation}.

\begin{table}[t]
\small
    \centering
    \begin{tabular}{c l}
    \toprule
      {\bfseries Symbol} & {\bfseries Description}\\
      \midrule
      $\mathcal{U}$ & The set of users in a recommender system\\
      $\mathcal{V}$ & The set of items in a recommender system\\
      $\mathcal{T}$ & The set of user-item interactions in a recommender system\\
      $\mathcal{F}$ & The set of features in a recommender system\\
      $\mathcal{S}$ & The set of sentiments in a recommender system\\
      $m$ & The number of users\\
      $n$ & The number of items\\
      $r$ & The number of features\\
      $u$ & A user ID in a recommender system\\
      $v$ & An item ID in a recommender system\\
      $f$ & A feature index in a recommender system\\
      $s$ & A sentiment index in a recommender system\\
      $\boldsymbol{A}$ & A user-feature attention matrix\\
      $\boldsymbol{B}$ & A item-feature quality matrix \\
      $\boldsymbol{A}^{cf}$ & The user-feature attention matrix after intervention with $\Delta_u$\\
      $\boldsymbol{B}^{cf}$ & The item-feature quality matrix after intervention with $\Delta_v$\\
      $\mathcal{G}_0$ & The set of popular items\\
      $\mathcal{G}_1$ & The set of long-tailed items\\
      $y_{uv}$ & Ground-truth value of the pair $(u, v)$\\
      $\hat{y}_{uv}$ & Predicted value of the pair $(u, v)$\\
      $K$ & The length of the recommendation list\\
      $\mathcal{R}_K$ & The set of recommendation lists with length K for all users\\
      $\Theta$ & Parameters of black-box recommendation model \\
      \bottomrule
    \end{tabular}
    \vspace{5pt}
    \caption{Summary of the notations in this work.}
    \label{Table:notation}
     \vspace{-10pt}
\end{table}


\subsection{Fairness and Disparity}
In this work, we consider explaining the exposure unfairness due to popularity bias in recommendation.
Given a recommendation model $g$, we will have a certain recommendation result $\mathcal{R}_K = \{\mathcal{R}(u_1,K), \mathcal{R}(u_2,K), \cdots, \mathcal{R}(u_m,K)\}$ containing all users' top-$K$ recommendation lists.
These recommendations determine the exposures of items, which is used to measure the fairness and disparity of the model.
We then split items into two groups based on their number of exposures in the recommendation list and denote $\mathcal{G}_0$ as popular item group and $\mathcal{G}_1$ as long-tailed item group.
Based on the above notations, we list some popular algorithmic fairness definitions related to popularity bias as follows:

\subsubsection{\textbf{Demographic Parity (DP)}}
Demographic parity in recommendation scenarios requires that the average exposure of the items from each group is equal~\cite{singh2018fairness,ge2021towards}. 
First, given $\mathcal{R}_K$, we denote the number of exposures in group $\mathcal{G}_{l}$ as
\begin{equation} \label{eq:expo}
\small
    \text {Exposure} \left(\mathcal{G}_{l} | \mathcal{R}_K  \right) = \sum_{u \in \mathcal{U}}  \sum_{v \in \mathcal{R}(u,K)} \mathcal{I}(v \in \mathcal{G}_l),l \in \{0,1\}.
\end{equation}
where $\mathcal{I}$ is the indicator function.

Then, we can express demographic parity fairness as follows,
\begin{equation}
\small
    \frac{\text {Exposure}\left(\mathcal{G}_{0} | \mathcal{R}_K \right)}{|\mathcal{G}_{0}|}=\frac{\text {Exposure}\left(\mathcal{G}_{1} | \mathcal{R}_K \right)}{|\mathcal{G}_{1}|},
\end{equation}
where groups $\mathcal{G}_0$ and $\mathcal{G}_1$ are created based on the item popularity, as mentioned before.

\subsubsection{\textbf{Exact-$K$ Fairness (EK)}}
Following \cite{ge2021towards}, we can also use the Exact-$K$ fairness in ranking, which requires the proportion/chance of protected candidates in every top-$K$ recommendation list remains statistically indistinguishable from a given maximum $\alpha$. 
This kind of fairness constraint is more suitable and feasible in practice for recommender systems as the system can adjust the value of $\alpha$. 
The concrete form of this fairness is shown as below,
\begin{equation} \label{eq:alpha_expo}
\small
    \frac{\text {Exposure}\left(\mathcal{G}_{0} | \mathcal{R}_K \right)}{\text {Exposure}\left(\mathcal{G}_{1} | \mathcal{R}_K \right)} = \alpha
\end{equation}
where $\alpha \in (0,1)$. Note that when $\alpha=\frac{|\mathcal{G}_0|}{|\mathcal{G}_1|}$ and the equation holds strictly, the above expression would be exactly the same as demographic parity.

\subsubsection{\textbf{Disparity}}
In practice, we can take the difference between the two sides of the equalities in the above definitions as a quantification measure for disparity. For example,
\begin{equation}
    \Psi_{DP} = |\mathcal{G}_{1}| \cdot \text {Exposure}\left(\mathcal{G}_{0} | \mathcal{R}_K \right) - |\mathcal{G}_{0}| \cdot \text {Exposure}\left(\mathcal{G}_{1} | \mathcal{R}_K \right)
\end{equation}

\begin{equation}
    \Psi_{EK} = \text {Exposure}\left(\mathcal{G}_{0} | \mathcal{R}_K \right) - \alpha \cdot \text {Exposure}\left(\mathcal{G}_{1} | \mathcal{R}_K \right)
\end{equation}
are two popular algorithm disparity measures used in fairness learning algorithms \cite{ge2021towards}.


\subsection{Counterfactual Reasoning}\label{sec:cef}
With the above notations and definitions of item exposure fairness, we can measure the disparity of the top-$K$ recommendation result $\mathcal{R}_K$. 
Then, the objective of our counterfactual reasoning problem is to generate feature-based explanations for the given black-box recommendation model $g$. 
The essential idea of the proposed explanation model is to discover a slight change $\Delta_v$ on each feature via solving a counterfactual optimization problem, which minimizes the disparity and a perturbation constraint that represents the effort to change the disparity, so that we can know which feature(s) are the underlying reasons for model disparity.

Specially, for each user-feature vector $\boldsymbol{A}_{:f}$, we slightly intervene with a vector $\Delta_u \in \mathrm{R}^m$ (and for each item-feature vector $\boldsymbol{B}_{:f}$, we intervene with $\Delta_v \in \mathrm{R}^n$), more specifically, the value of certain user feature $f$ for all users $\boldsymbol{A}_{:f}$ will be added to $\Delta_u$ and get $\mathbf{A}^{cf}$, (or the value of certain item feature $f$ for all items $\boldsymbol{B}_{:f}$ will be added to $\Delta_v$ and get $\boldsymbol{B}^{cf}$).
With the new user-feature matrix $\boldsymbol{A}^{cf}$ and item-feature matrix $\boldsymbol{B}^{cf}$, $g$ will change the recommendation result from $\mathcal{R}_K$ to a counterfactual result $\mathcal{R}_K^{cf}$.
More importantly, this will also change the fairness measure of that result to $\Psi^{cf}$, where $\Psi^{cf}$ can either be $\Psi_{EK}^{cf}$ or $\Psi_{DP}^{cf}$ depending on the choice of disparity.
And our goal is to look for the minimum intervention on user/item feature that is able to result in the greatest reduction in terms of disparity or unfairness.
Thus, objective function would be:
\begin{equation}\label{eq:obj1}
    \min \| \Psi^{cf} \|^2_2+ \lambda \|\Delta\|_2
\end{equation}
where $\Delta$ can be either $\Delta_u$ or $\Delta_v$ or the concatenation of them ($\Delta = [\Delta_u,\Delta_v]$), $\lambda \in (0,1)$ is a hyper-parameter that is used to control the weight between the two terms, and $\Psi^{cf}$ can be:
\begin{equation}\label{eq:psi_ek}
\small
\begin{aligned}
         \Psi_{EK}^{cf} &= \text{Exposure}(\mathcal{G}_0 | \mathcal{R}_K^{cf}) - \alpha \cdot \text{Exposure}(\mathcal{G}_1 | \mathcal{R}_K^{cf})\\
         &= \sum_{u \in \mathcal{U}}  \sum_{v \in \mathcal{R}^{cf}(u,K)} \mathcal{I}(v \in \mathcal{G}_0) - \alpha \cdot \sum_{u \in \mathcal{U}}  \sum_{v \in \mathcal{R}^{cf}(u,K)} \mathcal{I}(v \in \mathcal{G}_1)
\end{aligned}
\end{equation}
\begin{equation}\label{eq:psi_dp}
\small
\begin{aligned}
         \Psi_{DP}^{cf} &= \text{Exposure}(\mathcal{G}_0 | \mathcal{R}_K^{cf}) -  \frac{|\mathcal{G}_0|}{|\mathcal{G}_1|} \cdot \text{Exposure}(\mathcal{G}_1 | \mathcal{R}_K^{cf})\\
         &= \sum_{u \in \mathcal{U}}  \sum_{v \in \mathcal{R}^{cf}(u,K)} \mathcal{I}(v \in \mathcal{G}_0) - \frac{|\mathcal{G}_0|}{|\mathcal{G}_1|} \cdot \sum_{u \in \mathcal{U}}  \sum_{v \in \mathcal{R}^{cf}(u,K)} \mathcal{I}(v \in \mathcal{G}_1)
\end{aligned}
\end{equation}



A major challenge to optimize Eq. \eqref{eq:obj1} is the non-differentiable nature of $\Psi^{cf}$.
As a relaxation, we replace the indicator function $\mathcal{I}(\cdot)$ in the original definition (Eq. \eqref{eq:psi_dp} or Eq. \eqref{eq:psi_ek}) with $g(\cdot,\cdot)$, which is the predicted ranking score, and normalize the final results to stabilize the gradients of the objective function.
And the resulting disparity metric $\Tilde{\Psi}^{cf}$ becomes:
\begin{equation}
\small
    \Tilde{\Psi}_{EK}^{cf} = \frac{\sum_{u \in \mathcal{U}}  \left( \sum_{v \in \mathcal{G}_0 \cap \mathcal{R}^{cf}(u,K)} g(\boldsymbol{A}^{cf}_u,\boldsymbol{B}^{cf}_v) - \alpha \sum_{v \in \mathcal{G}_1 \cap \mathcal{R}^{cf}(u,K)} g(\boldsymbol{A}^{cf}_u,\boldsymbol{B}^{cf}_v) \right )}{\sum_{u \in \mathcal{U}}  \sum_{v \in \mathcal{R}^{cf}(u,K)} g(\boldsymbol{A}^{cf}_u,\boldsymbol{B}^{cf}_v)}
\end{equation}
or becomes:
\begin{equation}
\small
    \Tilde{\Psi}_{DP}^{cf} = \frac{\sum_{u \in \mathcal{U}}  \left( \sum_{v \in \mathcal{G}_0 \cap \mathcal{R}^{cf}(u,K)} g(\boldsymbol{A}^{cf}_u,\boldsymbol{B}^{cf}_v) - \frac{|\mathcal{G}_0|}{|\mathcal{G}_1|} \sum_{v \in \mathcal{G}_1 \cap \mathcal{R}^{cf}(u,K)} g(\boldsymbol{A}^{cf}_u,\boldsymbol{B}^{cf}_v) \right )}{\sum_{u \in \mathcal{U}}  \sum_{v \in \mathcal{R}^{cf}(u,K)} g(\boldsymbol{A}^{cf}_u,\boldsymbol{B}^{cf}_v)}
\end{equation}

Thus, our final objective for a given feature is 
\begin{equation}\label{eq:obj2}
    \min \|\Tilde{\Psi}^{cf} \|^2_2+ \lambda \|\Delta\|_2.
\end{equation}
The first term aims to realize the greatest reduction in terms of pre-defined disparity or unfairness.
The second perturbation constraint represents the edit distance between original inputs and the corresponding counterfactuals.
Finally, for each feature, we solve an optimization problem defined as Eq. \eqref{eq:obj2} and use the corresponding counterfactual recommendation result to calculate the explainability score, which will be detailed in the next section.

\subsection{Generate Feature-based Explanations} \label{sec:es}
For each feature in the feature space, we will solve the optimization problem defined as Eq. \eqref{eq:obj2} and consider $\Delta$ as the only trainable parameter. 
Once finished optimizing, we will get the ``minimial'' change $\Delta$ and the corresponding recommendation results under such change to that feature.
Then, we use \textit{Proximity}---the average edit distance between original input and the corresponding counterfactual---to measure the degree of perturbation.
And we use \textit{Validity}---the change of fairness caused by the feature's perturbation---to measure the degree of influence on fairness \cite{moraffah2020causal, verma2020counterfactual,mothilal2020explaining,tan2021counterfactual}.
\begin{equation}
\textit{Proximity} =  \|\Delta\|^2_2
\end{equation} 
\begin{equation}
\small
\begin{aligned}
    \textit{Validity} &= \frac{\text {Exposure}\left(\mathcal{G}_{0} | \mathcal{R}_K \right) -  \text {Exposure}\left(\mathcal{G}_{1} | \mathcal{R}_K \right)}{m \cdot K} \\
    &- \frac{\text {Exposure}\left(\mathcal{G}_{0} | \mathcal{R}^{cf}_K \right) -  \text {Exposure}\left(\mathcal{G}_{1} | \mathcal{R}^{cf}_K \right)}{m \cdot K},
\end{aligned}
\end{equation}
where $m$ is the number of users and $K$ is the length of recommendation lists.

Finally, the explainability score ($ES$) is the linear combination of \textit{Proximity} and \textit{Validity}, which is shown as follows:
\begin{equation}
    ES = \textit{Validity} - \beta \cdot \textit{Proximity},
\end{equation}
where $\beta \in (0,1)$ and larger score represents better explainability.

This score determines the ranking of a feature in terms of its ability to reduce the disparity of model $g$ while keeping the perturbation small.
Note that the original value of the feature corresponds to the optimal recommendation utility of $\mathcal{R}_K$ that the model $g$ learned, so larger proximity score may imply a greater sacrifice of utility.
Thus, the inclusion of this term in the objective function and the scoring function will result in an explanation finding process that is aware of the influence on both the fairness and recommendation utility.


\section{Experiments}

\subsection{Datasets}
To evaluate the models under different data scales, data sparsity and application scenarios, we perform experiments on three widely-used real-world datasets \cite{he2016ups, geng2022recommendation,ge2019maximizing,tan2021counterfactual,li2022autolossgen}.
Some basic statistics of the experimental datasets are shown in Table \ref{tab:dataset}.

\begin{itemize}
    \item \textbf{Yelp} dataset\footnote{\url{https://www.yelp.com/dataset}} contains users' reviews on various kinds of businesses such as restaurants, dentists, salons, etc. 
    This dataset contains 6,685,900 reviews, 192,609 businesses, 200,000 pictures in 10 metropolitan areas.
    
    \item \textbf{Amazon} dataset contains user reviews on products in Amazon e-commerce system\footnote{\url{https://nijianmo.github.io/amazon/index.html}}.
    The Amazon dataset contains 29 sub-datasets corresponding to 29 product categories. 
    We adopt two datasets of different scales to evaluate our method, which are \textit{CDs \& Vinyl} and \textit{Electronics}. 
\end{itemize}

Since the Yelp and Amazon review datasets are very sparse, similar as previous work \cite{zhangsigir14,Wang2018,tan2021counterfactual}, we remove the users and items with fewer than 20 reviews.
For each dataset, we first sort the records of each user based on the timestamp, and then hold-out the last 5 interacted items together with 100 randomly sampled negative items for each user to serve as the test data to evaluate black-box recommenders and do fairness explanation. 
The last item in the training set of each user is put into the validation set.
Since we focus on item exposure fairness, we need to split items into two groups $\mathcal{G}_0$ and $\mathcal{G}_1$ based on item popularity.
It would be desirable if we have the item impression/listing information and use it to group items, however, since Yelp and Amazon datasets are public dataset and only have interaction data, we use the number of interaction to group items in them.
Specifically, for Yelp and Amazon review datasets, the top 20\% items in terms of number of interactions belong to the popular group $\mathcal{G}_0$, and the remaining 80\% belong to the long-tail group $\mathcal{G}_1$. 

\begin{table}[h]
\caption{Basic statistics of the experimental datasets.}
\label{tab:dataset}
\centering
\setlength{\tabcolsep}{5pt}
\begin{adjustbox}{max width=\linewidth}
\begin{tabular}
    {lccccc} \toprule
    Dataset       & \#User & \#Item & \#Review & \#Aspect & Density \\ \hline
Yelp          & 12,028 & 20,181 & 502,158       & 106       & 0.208\%         \\
CDs \& Vinyl & 3,225       & 46,709       & 179,992    &  118          & 0.119\%            \\
Electronics    & 2,762       & 19,449    &    51,777         & 77           & 0.096\%              \\ 
\bottomrule
\end{tabular}
\end{adjustbox}
\vspace{-10pt}
\end{table}

\begin{figure*}[t]
\mbox{
\centering
    \subfigure[NDCG@5 vs Long-tail Rate@5 on Yelp]{
        \includegraphics[width=0.29\textwidth]{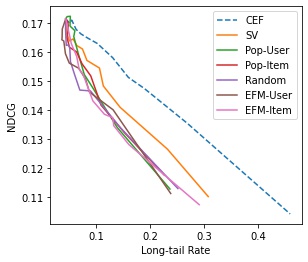}}
    \hspace{15pt}
    \subfigure[NDCG@5 vs Long-tail Rate@5 on Electronics]{
        \includegraphics[width=0.29\textwidth]{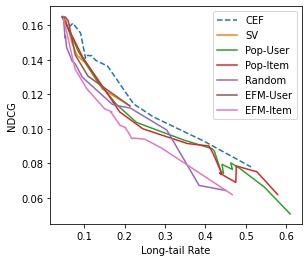}}
    \hspace{15pt}
    \subfigure[NDCG@5 vs Long-tail Rate@5 on CDs\&Vinyl]{
        \includegraphics[width=0.29\textwidth]{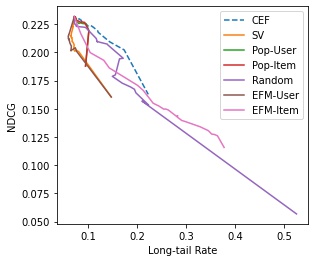}}
}
\mbox{
\centering
    \subfigure[NDCG@20 vs Long-tail Rate@20 on Yelp]{
        \includegraphics[width=0.29\textwidth]{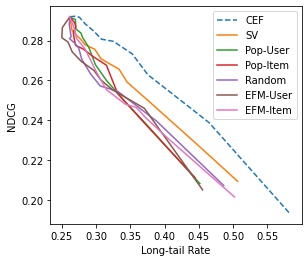}}
    \hspace{15pt}
    \subfigure[NDCG@20 vs Long-tail Rate@20 on Electronics]{
        \includegraphics[width=0.29\textwidth]{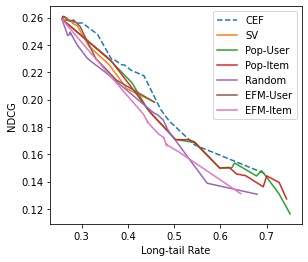}}
    \hspace{15pt}
    \subfigure[NDCG@20 vs Long-tail Rate@20 on CDs\&Vinyl]{
        \includegraphics[width=0.29\textwidth]{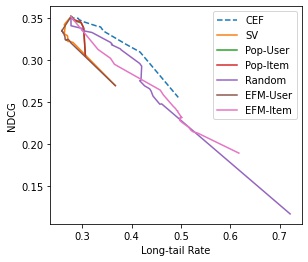}}
}
\mbox{
\centering
    \subfigure[NDCG@50 vs Long-tail Rate@50 on Yelp]{
        \includegraphics[width=0.29\textwidth]{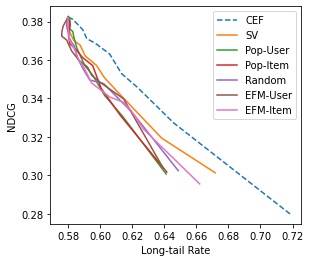}}
    \hspace{15pt}
    \subfigure[NDCG@50 vs Long-tail Rate@50 on Electronics]{
        \includegraphics[width=0.29\textwidth]{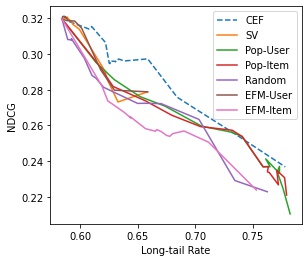}}
    \hspace{15pt}
    \subfigure[NDCG@50 vs Long-tail Rate@50 on CDs\&Vinyl]{
        \includegraphics[width=0.29\textwidth]{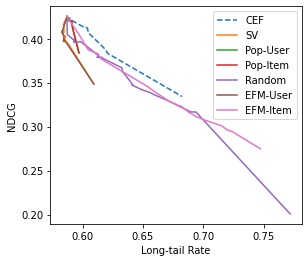}}
}
\caption{The accuracy-fairness trade-off curves for NDCG and Long-tail Rate on various datasets. The upper-right corner of each figure (high accuracy, low disparity) is preferred. Each data point is generated by cumulatively removing top 5 features in the explanation lists provided by explanation methods. }
\label{fig:trade-off}
\end{figure*}

\subsection{\textbf{Black-box Recommender System}}
As mentioned before, we first follow prior works \cite{zhangsigir14,chen2020try,Wang2018} to build a user-feature attention matrix and an item-feature quality matrix, and use both matrices together with the user-item interaction history to train a feature-aware recommendation model.

In this work, to demonstrate the idea of counterfactual explainable fairness, we use a simple deep neural network as the implementation of the recommendation model $g$, which includes one fusion layer followed by three fully connected layers with size \{$256$, $64$, $1$\}. 
The architecture of the fusion layer depends on how we are going to merge the user-feature and item-feature vectors (as is provided in Eq. \eqref{eq:product} and Eq. \eqref{eq:concatenation}).
Specifically, for Element-wise Product merge, the fusion layer is \{2 $\times$ feature size, 256\}, while for Concatenation merge, it is \{feature size, 256\}.
The final output layer is a sigmoid activation function so as to map $\hat{y}_{u,v}$ into the range of $(0, 1)$. 


The model parameters are optimized by stochastic gradient descent (SGD) optimizer with a learning rate of $0.01$. 
After the recommendation model is trained, all the parameters will be fixed in the counterfactual reasoning phase and explanation evaluation phase.
The recommendation performance on Element-wise product merge (Eq. \eqref{eq:product}) and Concatenation merge (Eq. \eqref{eq:concatenation}) are presented in Tab. \ref{tab:rec_result}.
For convenience and simplicity, the evaluations of fairness explanation methods presented in the experiment section are based on Element-wise product merge (Eq. \eqref{eq:product}).

\begin{table}
\centering
\begin{adjustbox}{max width=\linewidth}
\setlength{\tabcolsep}{7pt}
\begin{tabular}
    {m{2cm} c c c c} \toprule
    \multirow{2}{*}{Recommender} 
    & \multicolumn{2}{c}{F1 (\%)} 
    & \multicolumn{2}{c}{NDCG (\%)}\\
    \cmidrule(lr){2-3} \cmidrule(lr){4-5}
    & \multicolumn{1}{c}{@5 $\uparrow$} 
    & \multicolumn{1}{c}{@20 $\uparrow$} 
    & \multicolumn{1}{c}{@5 $\uparrow$} 
    & \multicolumn{1}{c}{@20 $\uparrow$}\\ \midrule
    \multicolumn{5}{c}{Yelp} \\\midrule
    Element-wise & 17.161 & 16.563 & 16.069 & 29.192\\
    Concatenation  & 16.266 & 16.929 & 17.338 & 29.780\\ \midrule
    \multicolumn{5}{c}{Electronics} \\\midrule
    Element-wise & 15.112 & 13.975 & 16.384 & 25.886\\
    Concatenation  & 15.083 & 14.044 & 16.350 & 25.946\\ \midrule
    \multicolumn{5}{c}{CDs \& Vinyl} \\\midrule
    Element-wise & 21.463 & 18.517 & 23.150 & 35.162\\
    Concatenation  & 20.737 & 18.443 & 22.393 & 34.672\\
\bottomrule
\end{tabular}
\end{adjustbox}
\vspace{10pt}
\caption{Summary of recommendation performance on three datasets for black-box recommendation models using Element-wise Product merge (Eq. \eqref{eq:product}) and Concatenation merge (Eq. \eqref{eq:concatenation}) in term of F1 and NDCG.}
\label{tab:rec_result}
\vspace{-10pt}
\end{table}

\subsection{Baselines}
Since there is no existing method specifically designed to explain fairness in recommendation. 
We adopt the following explanation methods as baselines:

\begin{itemize}
    \item \textbf{Random}: We randomly choose multiple features from the feature space without replacement and use them as explanation results. 
    
    \item \textbf{Popularity}: We rank all the features in the user-feature matrix and item-feature matrix based on their number of existences, and select the top ones as explanations, and denote them as \textbf{Pop-User} and \textbf{Pop-Item}, respectively.
    
    \item \textbf{EFM} \cite{zhangsigir14}: The Explicit Factor Model (EFM) for explainable recommendation. This work integrates matrix factorization with explicit features to align latent factors with explicit aspects for explanation. In this way, it predicts the user-feature preference scores and item-feature quality scores. The orgianl EFM uses the element-wise product of user-feature vector and item-feature vector and select the top ones as explanations to a given user-item pair. To generate global explanations, we calculate the average value of each feature from both user side and item side, and use them as explanations. Therefore, we have \textbf{EFM-User} and \textbf{EFM-Item}. Note that these features only explains the recommendation utility but do not explain the fairness.
    
    
    \item \textbf{Feature-based Explanation by Shapley Values (SV)}: \citeauthor{begley2020explainability} \cite{begley2020explainability} leveraged Shapley value-based methods to attribute the model disparity as the sum of individual contributions from input features to understand which feature contributes more or less to the model disparity.
    Considering the large number of features in the feature space, instead of using all possible coalitions, which is $r!$, we randomly sample 100 feature coalitions to calculate the Shapley value for each feature.
\end{itemize}

For \textbf{CEF}, we choose to minimize Eq. \eqref{eq:obj2}, where $\Delta = [\Delta_u,\Delta_v]$, $\Psi^{cf} = \Psi_{DP}^{cf}$ (Eq. \eqref{eq:psi_dp}), and $\frac{|\mathcal{G}_0|}{|\mathcal{G}_1|}=\frac{1}{4}$.
We set the hyper-parameter $\lambda = 1$ and $K = 5$. 
The model parameters are optimized by Adam optimizer with a learning rate of 0.01.

\begin{table*}[]
\caption{Summary of the performance and fairness on three datasets. 
We evaluate for ranking ($F1$ and $NDCG$, in percentage (\%) values, \% symbol is omitted in the table for clarity) and fairness ($KL\ Divergence$ and $Long-tail\ Rate$, also in \% values) with top 5 recommended items, and $E$ is the number of erased features.
Bold scores are used to indicate the greatest values.
}
\centering
\begin{adjustbox}{max width=\linewidth}
\setlength{\tabcolsep}{7pt}
\begin{tabular}
    {m{1.53cm} ccc ccc ccc ccc} \toprule
    \multirow{2}{*}{Methods} 
    & \multicolumn{3}{c}{F1@5(\%) $\uparrow$} 
    & \multicolumn{3}{c}{NDCG@5 (\%) $\uparrow$} 
    & \multicolumn{3}{c}{Long-tail Rate@5 (\%) $\uparrow$}
    & \multicolumn{3}{c}{KL@5 (\%) $\downarrow$}\\\cmidrule(lr){2-4} \cmidrule(lr){5-7} \cmidrule(lr){8-10} \cmidrule(lr){11-13}
 & E=5 & E=10 & E=20 & E=5 & E=10 & E=20 & E=5 & E=10 & E=20 & E=5 & E=10 & E=20 \\\midrule 

\multicolumn{13}{c}{Yelp} \\\midrule
Random & 15.671 & 15.345 & 14.809 & 16.788 & 16.255 & 15.674 & 4.3191 & 4.8066 & 5.2302 & 10.506 & 9.6994 & 9.0410 \\
Pop-User & \textbf{16.074} & \textbf{15.636} & 14.956 & 17.236 & \textbf{16.748} & 15.983 & 4.6047 & 6.0428 & 6.6289 & 10.026 & 7.8770 & 7.1107 \\
Pop-Item & 16.050 & 15.498 & 14.868 & \textbf{17.055} & 16.522 & 16.013 & 4.7180 & 4.5703 & 6.3580 & 9.8421 & 10.083 & 7.4577 \\
EFM-User & 15.735 & 15.370 & 14.710 & 16.804 & 16.392 & 15.626 & 3.7084 & 3.9940 & 5.0086 & 11.598 & 11.075 & 9.3808 \\
EFM-Item & 15.538 & 14.434 & 13.533 & 16.558 & 15.406 & 14.320 & \textbf{5.0874} & 6.8406 & 9.3622 & \textbf{9.2588} & 6.8477 & 4.2092 \\
SV & 15.680 & 15.188 & 14.814 & 16.700 & 16.235 & 15.719 & 4.4570 & 6.3974 & 8.2688 & 10.272 & 7.4064 & 5.2486 \\
CEF & 15.897 & 15.513 & \textbf{15.296} & 17.015 & 16.635 & \textbf{16.309} & 5.0233 & \textbf{7.1706} & \textbf{10.169} & 9.3579 & \textbf{6.4518} & \textbf{3.5328} \\
\midrule

\multicolumn{13}{c}{Electronics} \\\midrule
Random & 14.981 & 14.960 & 14.945 & 15.253 & 15.272 & 15.336 & 5.2715 & 5.4018 & 5.3439 & 8.9788 & 8.7846 & 8.8705 \\
Pop-User & 13.330 & 11.940 & 9.9782 & 14.417 & 12.795 & 10.361 & 8.6676 & 13.164 & 22.947 & 4.8522 & 1.6141 & \textbf{0.2621} \\
Pop-Item & 13.149 & 11.701 & 9.6017 & 14.118 & 12.482 & 9.9908 & \textbf{9.6886} & \textbf{14.460} & \textbf{24.540} & \textbf{3.9269} & \textbf{1.0372} & 0.6115 \\
EFM-User & \textbf{15.018} & 15.018 & 15.018 & \textbf{16.454} & \textbf{16.451} & 16.426 & 4.6125 & 4.4822 & 4.7139 & 10.014 & 10.230 & 9.8487 \\
EFM-Item & 12.541 & 11.622 & 10.586 & 13.453 & 12.314 & 10.976 & 7.7552 & 10.644 & 16.857 & 5.7903 & 3.1690 & 0.3218 \\
SV & 15.061 & \textbf{15.126} & \textbf{15.112} & 16.379 & 16.418 & \textbf{16.487} & 5.0615 & 4.9312 & 4.9674 & 9.2987 & 9.5019 & 9.4451 \\
CEF & 14.829 & 14.887 & 13.164 & 15.956 & 16.115 & 14.149 & 6.5821 & 7.1976 & 10.275 & 7.1697 & 6.4201 & 3.4500 \\
\midrule

\multicolumn{13}{c}{CDs \& Vinyl} \\\midrule
Random & \textbf{21.463} & 21.246 & 21.103 & 22.131 & 21.968 & 21.821 & 7.2062 & 7.3612 & 7.5906 & 6.4102 & 6.2307 & 5.9715 \\
Pop-User & 21.413 & 21.432 & \textbf{21.432} & 23.118 & 23.133 & \textbf{23.162} & 7.1937 & 7.1999 & 7.2496 & 6.4247 & 6.4174 & 6.3596 \\
Pop-Item & 21.457 & \textbf{21.469} & 21.413 & \textbf{23.156} & \textbf{23.196} & 23.150 & 7.2062 & 7.2062 & 7.2186 & 6.4102 & 6.4102 & 6.3957 \\
EFM-User & 20.241 & 20.210 & 20.055 & 21.621 & 21.594 & 21.482 & 6.1395 & 6.0651 & 6.0093 & 7.7465 & 7.8468 & 7.9226 \\
EFM-Item & 19.968 & 18.381 & 17.159 & 21.653 & 19.972 & 18.619 & \textbf{8.5271} & \textbf{10.449} & \textbf{14.325} & \textbf{4.9896} & \textbf{3.3154} & \textbf{1.0908} \\
SV & 20.675 & 20.700 & 20.545 & 22.290 & 22.283 & 22.174 & 6.9271 & 6.8403 & 6.9147 & 6.7424 & 6.8481 & 6.7574 \\
CEF & \textbf{21.463} & 21.438 & 21.333 & 23.099 & 23.061 & 22.962 & 7.2124 & 7.2496 & 7.4046 & 6.4029 & 6.3596 & 6.1811 \\
\bottomrule
\end{tabular}\label{tab:result}
\end{adjustbox}
\end{table*}

\subsection{Evaluation Methods and Metrics}
Once we obtain the feature-based explanations from each baseline as well as our proposed CEF, we need to compare the effectiveness of these results, in other words, their contributions to the fairness-utility trade-off.
In order to evaluate the feature-based explanations, we follow the widely deployed erasure-based evaluation criterion in Explainable AI.
The intuition behind the erasure-based criterion is to measure how much the model performance would drop after the set of the ``most important'' features in an explanation is removed~\cite{zaidan2007using,yu2019rethinking}.
Similarly, in the setting of explainable fairness, we use it to measure the fairness-utility trade-off in recommendation, namely, how much the recommendation performance would drop and how much the recommendation fairness would improve after the set of the ``most important'' features in an explanation is removed.
Specifically, for each feature-based explanation result, we erase the set of the ``most important'' features in both the user-feature and item-feature matrices for all users and items, then input the erased user-feature and item-feature matrices into pre-trained recommendation model $g$ to generated a new recommendation results.
Based on the recommendation performance and fairness of the new results, we compare the effectiveness of each explainable fairness methods.

We select several most commonly used top-$K$ ranking metrics to evaluate the model's recommendation performance after erasure, including \textbf{F1 Score}, and \textbf{NDCG}.
For fairness evaluation, we define \textbf{Long-tail Rate}, which simply refers to the ratio of the number of long-tailed items in the recommendation list to the total number of items in the list.
We also employ \textbf{KL-divergence} (KL) to compute the expectation of the difference between protected group membership at top-$K$ vs. in the overall population, which is:
\begin{equation}
    d_{K L}\left(D_{1}|| D_{2}\right)=\sum_{j\in\{0,1\}} D_{1}(j) \ln \frac{D_{1}(j)}{D_{2}(j)}
\end{equation}
where $D_1$ represents the true group distribution between $\mathcal{G}_0$ and $\mathcal{G}_1$ in top-$K$ recommendation list, and $D_2=[\frac{|\mathcal{G}_0|}{|\mathcal{V}|},\frac{|\mathcal{G}_1|}{|\mathcal{V}|}]$ represents their ideal distribution of the overall population.


\subsection{Experimental Results}
The major experimental results are shown in Fig. \ref{fig:trade-off}, where we plot the fairness-utility trade-off, i.e., the relationship between NDCG and Long-tail Rate (namely, 1-Popularity Rate) with different length of recommendation lists (@5, @20, @50).
Since the relationship between F1 and and Long-tail Rate has very similar conclusions, we choose not to present them here.
Each data point Fig. \ref{fig:trade-off} is generated by cumulatively removing top 5 features in the remaining explanation list provided by each explanation method.  
We also present the values of F1@5, NDCG@5, Long-tail Rate@5 and KL@5 in Tab. \ref{tab:result} after removing top-5, top-10, top-20 features in each explanation result to quantitatively analyse the results.

First, in Fig. \ref{fig:trade-off} and Tab. \ref{tab:result}, we can easily find that all the methods, even randomly selecting features and erasing them, can improve recommendation fairness.
Besides, the higher the number of features we erase, the lower the disparity rate we can achieve.
This is easy to understand as erasing features will mitigate the representation gap between popular items and long-tailed items, causing more under-represented items to be recommended.
However, it also brings huge decline to the recommendation performance.
For example, compared with the original recommendation performance on NDCG@5, the method with the worst trade-off behavior drops relatively 2.119 \% on Yelp,  21.786 \% on Electronics, and 7.072 \% on CDs \& Vinly when deleting top 5 features.
Second, we can see that even though the idea of using popular features as explanations is very intuitive, their performance may even be worse than random selection, which indicates that compared with fairness, popular features either from user side or item side are more sensitive to recommendation performance, while random selection guarantees low probabilities of choosing those scarce features, which in turn results in better trade-off.
Third, the performances of SV are much worse than CEF as it only explains disparity alone, ignoring the inherent trade-off between fairness and utility.
Finally, in Fig. \ref{fig:trade-off}, where the blue dotted line represents the performance of our proposed CEF framework, it is obvious that the feature-based explanations provided by CEF are capable of achieving much better fairness-utility trade-off on datasets with various scales and densities.
Specifically, compared with the original recommendation performance on NDCG@5, CEF method drops only relatively 0.851 \% on Yelp, 2.682 \% on Electronics, and 0.594 \% on CDs $\&$ Vinly, while it increases relatively 13.431 \% on Yelp,  25.73 \% on Electronics, and on 3.085 \% CDs $\&$ Vinly at Long-tail Rate@5.


\begin{figure}[h]
  \centering
  \includegraphics[width=0.3\textwidth]{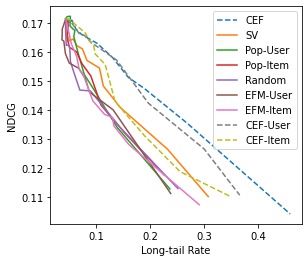}
  \caption{Ablation study on Yelp dataset.}
  \label{fig:ablation}
  \vspace{-15pt}
\end{figure}


\subsection{Ablation Studies}
As mentioned in Sec. \ref{sec:cef}, the choice of $\Delta$ in the objective function (Eq. \eqref{eq:obj1} or Eq. \eqref{eq:obj2}) can either be $\Delta_u$ or $\Delta_v$ or both of them, depending on how we are going to intervene the given feature in the feature space.
Besides, all the experimental results in Table \ref{tab:result} and Fig. \ref{fig:trade-off} are based on intervening the given feature using both $\Delta_u$ and $\Delta_v$ (namely, $\Delta = [ \Delta_u, \Delta_v ]$).
Therefore, to study how the choice of $\Delta$ is going to influence the experimental results, we run additional experiments based on the variants of the original CEF by either choosing $\Delta_u$ or $\Delta_v$ alone, denoted as CEF-User and CEF-Item, respectively.
The objective of CEF-User is 
$\min \|\Tilde{\Psi}_{EK}^{cf} \|^2_2+ \lambda \|\Delta_u\|_2.$
And that of CEF-Item is
$\min \|\Tilde{\Psi}_{EK}^{cf} \|^2_2+ \lambda \|\Delta_v\|_2.$
For convenience, we only present the results in Yelp dataset, as is shown in Fig. \ref{fig:ablation}. 
Similar conclusions are also achieved on other datasets.

As is shown in Fig. \ref{fig:ablation}, the evaluations on CEF-User and CEF-Item achieve worse fairness-utility trade-off when compared with the original CEF. 
This is understandable as CEF uses both $\Delta_u$ and $\Delta_v$ as its parameters, which is a much larger parameter space and can achieve better representations.
Moreover, even though CEF-User and CEF-Item are worse than CEF, they are still far more better than most of the baselines, especially, CEF-User is better than all the baselines, which indicates the effectiveness of our proposed framework.

\section{Conclusion and Future Work}
In this paper, we study the problem of explainable fairness in recommendation and propose a framework based on counterfactual reasoning, called CEF. To the best of our knowledge, this is the first work to introduce the idea of explainable fairness in recommender systems.
We design a learning-based counterfactual reasoning method to discover critical features that will significantly influence the fairness-utility trade-off and use them as fairness explanations for black-box feature-aware recommendation systems.
Extensive experiments have been conducted to evaluate the effectiveness of our proposed framework and the explanations generated by CEF can achieve better fairness-utility trade-off than all the baselines when using them to do fair learning.
In the future, we hope to design algorithmic methods that can generate multiple explanations at the same time without greedy choosing them through explainability scores. (One possible solution would be using penalizing vectors to control the number of perturbed features.)

\section*{Acknowledgments}
We appreciate the valuable feedback of the reviewers. 
This work was supported in part by NSF IIS 1910154, 2007907, 2046457 and Facebook Faculty Research Award. Any opinions, findings, conclusions or recommendations expressed in this material are those of the authors and do not necessarily reflect those of the sponsors.

\appendix


\section{Appendix: Case Study}
In this section, we provide the top-5 feature-based explanations that are generated by each method on Yelp dataset.
The explanation results are shown in Tab. \ref{tab:explanations}, which exactly verifies our motivation that it is difficult to manually identify feature explanations for exposure unfairness and popularity bias in recommender system. 
For example, it is hard to tell how input features (like chicken, cheese, pizza) would influence the exposure opportunity in restaurant recommendation.
Thus, we do need explainable fairness methods to identify such features in recommendation. 

\begin{table}[htbp]
    \centering
    \begin{tabular}{c l}
    \toprule
      {\bfseries Method} & {\bfseries Feature-based Explanations}\\
      \midrule
      Pop-User & food, service, chicken, prices, hour\\
      Pop-Item & food, service, prices, visit, hour\\
      EFM-User & store, patio, dishes, dish, rice\\
      EFM-Item & flavor, decor, dishes, inside, cheese\\
      SV & server, size, pizza, food, restaurant\\
      CEF & meal, cheese, dish, chicken, taste\\
      \bottomrule
    \end{tabular}
    \vspace{10pt}
    \caption{Top-5 feature-based explanations on Yelp dataset. }
    \label{tab:explanations}
\end{table}





\bibliographystyle{ACM-Reference-Format}
\bibliography{sample-base}


\end{document}